\documentclass[10pt,journal]{IEEEtran}

\usepackage{cmap}
\input{glyphtounicode}
\pdfglyphtounicode{Ccaron}{010C}
\pdfglyphtounicode{ccaron}{010D}
\pdfglyphtounicode{edotaccent}{0117}
\pdfglyphtounicode{scaron}{0161}

\usepackage[L7x,T1]{fontenc}
\usepackage[utf8]{inputenc}
\usepackage{amsmath,amssymb}
\usepackage{xcolor}
\usepackage{booktabs}
\usepackage{colortbl}
\usepackage{url}
\usepackage{ragged2e}
\usepackage{hyperref}
\usepackage{tikz}
\usetikzlibrary{shapes.geometric, arrows.meta, positioning, calc, fit, backgrounds}

\hypersetup{
    colorlinks=true,
    linkcolor=black,
    citecolor=black,
    urlcolor=blue
}

\usepackage{newunicodechar}
\newcommand{\inr}{\text{Rs.\ }}
\newunicodechar{₹}{\inr}

\DeclareRobustCommand{\CernevicieneName}{{\fontencoding{L7x}\selectfont Černevičienė}}
\DeclareRobustCommand{\KabasinskasName}{{\fontencoding{L7x}\selectfont Kabašinskas}}

\definecolor{bwBlack}{HTML}{000000}       % Pure Black
\definecolor{bwDark}{HTML}{1E293B}        % Charcoal Slate
\definecolor{bwMid}{HTML}{475569}         % Mid Slate
\definecolor{bwLight}{HTML}{F1F5F9}       % Soft Light Grey
\definecolor{bwWhite}{HTML}{FFFFFF}       % Clean White
\definecolor{bwAlt}{HTML}{E2E8F0}         % Container Fill

\tikzset{
    flowNode/.style={
        rectangle, rounded corners=2pt, draw=bwBlack, fill=bwWhite, line width=0.7pt,
        text=bwBlack, align=center, font=\scriptsize, inner sep=3pt, minimum height=0.48cm
    },
    flowDecision/.style={
        diamond, aspect=2, draw=bwBlack, fill=bwLight, line width=0.7pt,
        text=bwBlack, align=center, font=\scriptsize, inner sep=1.5pt
    },
    flowTerminal/.style={
        rectangle, rounded corners=4pt, draw=bwBlack, fill=bwDark,
        text=bwWhite, font=\scriptsize\bfseries, align=center, inner sep=3.5pt
    },
    flowAllow/.style={
        rectangle, rounded corners=2pt, draw=bwBlack, fill=bwAlt, line width=0.7pt,
        text=bwBlack, font=\scriptsize\bfseries, align=center, inner sep=3pt
    },
    flowBlock/.style={
        rectangle, rounded corners=2pt, draw=bwBlack, fill=bwLight, line width=0.9pt,
        text=bwBlack, font=\scriptsize\bfseries, align=center, inner sep=3pt, dashed
    },
    line/.style={
        -{Stealth[scale=0.75]}, draw=bwBlack, line width=0.6pt
    },
    dashedLine/.style={
        -{Stealth[scale=0.75]}, draw=bwBlack, line width=0.6pt, dashed
    },
    erTable/.style={
        rectangle, draw=bwBlack, line width=0.7pt, rounded corners=1pt,
        fill=bwWhite, inner sep=0pt, font=\tiny, align=left
    }
}

\makeatletter
\def\@IEEEcompsocleadtitleabstractindextext{}

\makeatother

\renewenvironment{abstract}{%
    \par\vspace{0.5em}%
    \noindent\textbf{\large Abstract}\par\vspace{0.35em}%
    \justifying\normalsize%
}{%
    \par\vspace{0.7em}%
}

\begin{document}

\title{\LARGE\textbf{FST Pay: Deterministic Safety-Gated Architecture\\for Youth Digital Payments}}

\author{
    \IEEEauthorblockN{\textbf{Shaikh Mohammed Burhan}, \textbf{Syed Farhaan Quadri}, and \textbf{Dr. Tabassum Nahid Sultana}} \\[0.35em]
    \IEEEauthorblockA{Department of Computer Science and Engineering (CSE) \\
    Khaja Bandanawaz University (KBNU), Kalaburagi, Karnataka 585104, India \\
    Email: burhan.mulla21@gmail.com, syedfarhaanquadri@gmail.com, tabassumcse85@gmail.com}
}

\maketitle

% =========================================================================
% ABSTRACT
% =========================================================================
\begin{abstract}
Digital payment infrastructures increasingly provide adolescent users with direct access to real-time financial services. While early access promotes financial literacy and digital inclusion, it exposes young users to severe risks of impulsive spending, social engineering frauds, unauthorized transactions, and merchant exploitation. Conventional countermeasures rely on probabilistic machine learning or rigid static controls. However, allowing probabilistic or generative artificial intelligence (AI) models to directly influence real-time payment authorization introduces non-determinism, unpredictable edge-case behavior, and critical audit vulnerabilities. This paper introduces Financial Safety for Teens Pay (FST Pay) as an architectural and formal specification; empirical validation is scoped for future testbed implementations. FST Pay is founded on an immutable operational boundary: strict deterministic safety gating on the real-time authorization path coupled with decoupled downstream AI explanation. Transactions initiated via interoperable rails such as the Unified Payments Interface (UPI) are subjected to six deterministic invariant checks covering spending limits, guardian co-sign policies, transaction amount thresholds, merchant category codes, temporal access intervals, and hardware/device integrity constraints. Transactions are classified strictly into ALLOW, REVIEW, or BLOCK outcomes without probabilistic ambiguity through an ordered, mutually exclusive decision function. High-risk transactions trigger an asynchronous guardian co-sign workflow. Generative AI is relegated entirely downstream of settlement, consuming published post-decision events solely to generate natural-language financial literacy insights and risk rationales without holding mutation privileges over the ledger. We formally specify the safety invariants, present end-to-end TikZ architectural models, and outline an auditable relational data schema.
\end{abstract}


\section{Introduction}
\IEEEPARstart{R}{eal-time} account-to-account payment systems have reshaped retail commerce worldwide~\cite{oecd2020advancing, basavesh2024upi}. In particular, open digital infrastructures such as India's Unified Payments Interface (UPI), developed by the National Payments Corporation of India (NPCI) and regulated by the Reserve Bank of India, process billions of high-velocity, low-cost financial transactions monthly~\cite{kukreja2024evidence, habibpour2023uncertainty}. The rapid democratization of mobile payment devices has lowered the entry age for digital transactions, granting adolescent demographics direct access to electronic fiat currency~\cite{oecd2020advancing}.

Extending uninhibited transactional autonomy to young, financially inexperienced users exposes them to asymmetric risks. Adolescents routinely exhibit vulnerability to impulsive overspending, deceptive gamification patterns (such as loot boxes and in-app microtransactions), phishing scams, and coercive social engineering~\cite{kukreja2024evidence, maslennikov2021risks}. Conventional retail banking architectures typically approach account safety by treating users as fully autonomous legal adults, enforcing standard multi-factor authentication (MFA) but lacking fine-grained parental oversight mechanisms~\cite{thomas2024family, choong2024childmfa}. In contrast, custodial bank accounts often impose total lockouts that impede autonomous learning for minor dependents~\cite{collins2024minors, zhenhe2024childbank}.

Modern payment networks employ advanced machine learning (ML) classifiers to detect fraudulent activities~\cite{cheng2024gnn, lebichot2024catastrophic, pozzolo2024credit}. While supervised classifiers, graph neural networks, and anomaly detectors excel at macroscopic fraud identification, their probabilistic nature poses fundamental hazards when such classifiers are directly embedded as primary authorization arbiters for minors. An authorization failure driven by latent weight drift or opaque statistical boundaries lacks the auditability and predictable guarantees essential in legal guardianship~\cite{nohr2024limits}. The National Institute of Standards and Technology Generative AI Profile (NIST AI 600-1)~\cite{nist2024genai} notes that foundation models suffer from stochastic hallucinations, prompt injection vulnerabilities, and non-deterministic decision paths~\cite{closer2024genai}. Allowing a generative model or an uncalibrated probabilistic classifier to directly trigger, modify, or decline a financial payment violates foundational financial safety tenets and model risk governance principles~\cite{nohr2024limits, mrm2024genai}.

Explainable artificial intelligence (XAI) has emerged as an indispensable requirement in financial technologies~\cite{cerneviciene2024xai}. When an automated system denies an adolescent's payment or suspends an account, merely presenting an obscure error code yields zero educational value and induces frustration~\cite{cerneviciene2024xai}. Parents and adolescent users require clear, natural-language rationales detailing why a transaction was restricted and how to establish safer spending patterns~\cite{wang2024digitalreg, cecilia2024familycomm}.

To resolve the tension between mathematical determinism in financial clearance and conversational adaptability in user explanation, this paper presents Financial Safety for Teens Pay (FST Pay) as an architectural and formal specification. The core philosophical premise dictates that zero probabilistic or generative components may exist on the synchronous authorization rail. Instead, every transaction request is passed through a deterministic verification engine that evaluates explicit, parentally configured safety invariants across multi-dimensional parameters. Only after an immutable state has been permanently committed to the ledger does an isolated downstream generative AI engine consume the audit event to compose plain-language educational summaries.

The main contributions of this paper are summarized as follows:
\begin{enumerate}
    \item We design and formalize a deterministic safety-gating architecture that eliminates probabilistic latency and hallucination risks from the real-time financial authorization path.
    \item We establish a dual-state guardian co-sign mechanism that cleanly intercepts high-risk or threshold-exceeding transactions and routes them into a synchronous multi-party consent hold.
    \item We formulate a strict non-interference invariant, enforced architecturally through capability isolation, preventing post-decision generative AI models from mutating ledger states.
    \item We detail a complete systems architecture, an end-to-end Unified Modeling Language (UML) interaction pipeline, a multi-participant sequence flow, and a fully normalized relational schema optimized for auditable youth payment systems.
\end{enumerate}

\section{Related Work}
\subsection{Youth Financial Inclusion and Payment Rails}
The digital transformation of retail banking has catalyzed research into the financial socialization of youth. The Organisation for Economic Co-operation and Development (OECD) emphasizes that youth-focused financial inclusion must balance access with protective scaffolding, as minors routinely face unique behavioral and digital exploitation threats~\cite{oecd2020advancing}. Interoperable architectures such as UPI offer instant settlement across bank accounts~\cite{chalise2024engineering}. While UPI incorporates end-to-end cryptographic signatures and device-binding protocols, it delegates user-level risk limits and behavioral spending rules to participating payment service provider (PSP) applications. Most commercial banking applications enforce coarse account-level caps rather than context-aware parental control policies~\cite{chalise2024engineering}. Commercial youth fintech offerings (such as Greenlight and FamPay) have popularized prepaid family allowances, but remain bound to proprietary static ledger limits without verifiable formal invariant guarantees or decoupled explanatory artificial intelligence.

\subsection{Machine Learning in Financial Fraud Detection}
Financial transaction environments present severe class imbalance, extreme throughput requirements, and evolving adversarial tactics~\cite{lebichot2024catastrophic, pozzolo2024credit}. Contemporary research categorizes machine learning models for payment fraud, noting that while deep learning and neural network architectures achieve high Area Under the Receiver Operating Characteristic Curve (AUROC), they require significant computational overhead and exhibit vulnerability to concept drift~\cite{cheng2024gnn, habibpour2023uncertainty}. Furthermore, probabilistic systems struggle to provide the hard mathematical guarantees required for statutory compliance and contractual policy enforcement.

\subsection{Explainable and Generative AI in Financial Systems}
Explainability in financial algorithms is vital for both regulatory compliance and user trust~\cite{cerneviciene2024xai}. Prior literature on explainable artificial intelligence (XAI) across financial information systems highlights methods like SHapley Additive exPlanations (SHAP) and Local Interpretable Model-agnostic Explanations (LIME) used to unpack post-hoc credit and fraud models. \CernevicieneName\ and \KabasinskasName~\cite{cerneviciene2024xai} synthesized XAI taxonomies, arguing that transparent, interpretable outputs are foundational for user-facing automated decision systems. With the emergence of Large Language Models (LLMs), recent research has explored conversational agents for customer interaction and contextual spending analytics~\cite{bauer2024expl, hallucination2024fdm}. However, as cataloged in the NIST Generative AI Risk Management Profile (NIST AI 600-1)~\cite{nist2024genai}, foundation models exhibit non-negligible failure modes including factual hallucinations, prompt injection vulnerabilities, and non-deterministic output paths.

\subsection{Research Gap and Architectural Opportunity}
Existing financial platforms typically commit to one of two suboptimal paradigms: either they deploy legacy static rules that lack contextual explanation, or they experiment with end-to-end neural pipelines that compromise auditability and latency guarantees~\cite{mlnn2024finance}. No unified framework explicitly addresses youth digital safety by synchronizing deterministic policy validation, real-time guardian co-signing, and asynchronous, read-only generative explanation. FST Pay is architected specifically to address this gap.

\section{Methodology}
\subsection{Foundational Baseline and Model Evolution}
The mathematical formalization underpinning FST Pay builds upon classical Attribute-Based Access Control (ABAC) theory and deterministic finite-state transaction machines~\cite{nohr2024limits, chandnani2024idempotency}. In standard access control paradigms, an authorization request is evaluated against a static Boolean predicate over subject, object, and environment attributes. However, digital youth payments introduce dynamic financial dependencies—such as rolling diurnal expenditures, fixed categorical Merchant Category Codes (MCC), diurnal curfew intervals, and real-time multi-party supervisory escalation—that standard access models cannot express.

To bridge this gap, we adapt and extend the generalized authorization tuple model into a six-dimensional invariant evaluation vector $\mathcal{I}$. Rather than permitting probabilistic scoring or unconstrained language model evaluation on the authorization path, our model maps each transaction request deterministically onto a closed ternary outcome space $\Omega = \{\text{ALLOW}, \text{REVIEW}, \text{BLOCK}\}$. Below, we define the constituent equations and explain how they advance traditional payment control logic.

\subsection{Mathematical Formalization of the Safety Invariant Model}
Let $\mathcal{T}_{\text{req}}$ denote the set of all valid transaction initiation requests. A specific payment initiation request $R \in \mathcal{T}_{\text{req}}$ from an adolescent user is formalized as an immutable tuple:
\begin{equation}
R = (u, m, a, t, d, c)
\end{equation}
where $u \in \mathcal{U}$ represents the authenticated adolescent user identity (e.g., \texttt{User\_2317}), $m \in \mathcal{M}$ denotes the counterparty payee identifier, $a \in \mathbb{R}^{+}$ denotes the transactional quantum in Indian Rupees (\inr INR), $t \in \mathcal{T}$ represents the timestamp, $d \in \mathcal{D}_{\text{dev}}$ denotes hardware device telemetry, and $c \in \mathcal{C}$ encapsulates contextual environment attributes.

Prior to invariant evaluation, we define the merchant universe partition. Let the global counterparty domain $\mathcal{M}$ be partitioned into three pairwise disjoint sets:
\begin{equation}
\mathcal{M} = \mathcal{M}_{\text{auto}} \cup \mathcal{M}_{\text{cosign}} \cup \mathcal{M}_{\text{prohibited}}
\end{equation}
where $\mathcal{M}_{\text{auto}} = \mathcal{M}_{\text{p2p\_whitelist}} \cup \mathcal{M}_{\text{preapproved\_retail}}$ denotes pre-authorized peer contacts and verified minor-safe merchants, $\mathcal{M}_{\text{cosign}}$ denotes unverified or high-value categories requiring supervisory approval, and $\mathcal{M}_{\text{prohibited}}$ represents blacklisted merchant categories (e.g., gambling or adult services). Furthermore, let $\theta_{\text{cosign}} \in \mathbb{R}^{+}$ represent the guardian-configured single-transaction monetary threshold above which autonomous execution is barred.

FST Pay evaluates request $R$ against six deterministic safety invariants $\mathcal{I} = \{I_{s}, I_{\text{auto}}, I_{a}, I_{m}, I_{t}, I_{r}\}$:

\subsubsection{Rolling Velocity Constraint ($I_s$)}
Uncontrolled fund drainage is prevented by tracking historical expenditure $\mathcal{H}_u(\Delta t)$ over a moving interval $\Delta t$, validating that the aggregate quantum respects the velocity cap $L_{\Delta t}$:
\begin{equation}
I_{s}(R) = \begin{cases}
1, & \text{if } a + \sum_{k \in \mathcal{H}_{u}(\Delta t)} a_{k} \le L_{\Delta t} \\
0, & \text{otherwise}
\end{cases}
\end{equation}
When breached ($I_s = 0$), the pipeline does not discard the request; it diverts the payment into supervisory review for explicit guardian clearance.

\subsubsection{Autonomous Clearance Invariant ($I_{\text{auto}}$)}
Low-risk transactions within established personal allowances and whitelisted networks should proceed without guardian latency:
\begin{equation}
I_{\text{auto}}(R) = \begin{cases}
1, & \text{if } m \in \mathcal{M}_{\text{auto}} \wedge a < \theta_{\text{cosign}} \\
0, & \text{otherwise (co-sign required)}
\end{cases}
\end{equation}
A value of $0$ indicates that the transaction involves an unfamiliar merchant, a non-whitelisted peer, or a monetary sum exceeding $\theta_{\text{cosign}}$, necessitating real-time parental consent.

\subsubsection{Single Transaction Quantum Ceiling ($I_a$)}
To defend the account against sudden balance exhaustion, a hard per-transaction ceiling $A_{\max}$ is enforced:
\begin{equation}
I_{a}(R) = \begin{cases}
1, & \text{if } a \le A_{\max} \\
0, & \text{otherwise}
\end{cases}
\end{equation}
In contrast to velocity limits, an overage here represents a parameter violation that halts the checkout pipeline immediately.

\subsubsection{Payee Category Restriction ($I_m$)}
Target merchant classification codes are cross-referenced with the prohibited partition $\mathcal{M}_{\text{prohibited}}$~\cite{westermeier2024financial}:
\begin{equation}
I_{m}(R) = \begin{cases}
1, & \text{if } \text{MCC}(m) \notin \mathcal{M}_{\text{prohibited}} \\
0, & \text{otherwise}
\end{cases}
\end{equation}
Interactions with restricted business sectors (e.g., MCC 7995) result in a non-negotiable block.

\subsubsection{Curfew Interval Enforcement ($I_t$)}
Transactions are confined to diurnal operating windows $[T_{\text{open}}, T_{\text{close}}]$, restricting transaction execution during unauthorized nighttime hours:
\begin{equation}
I_{t}(R) = \begin{cases}
1, & \text{if } \text{time}(t) \in [T_{\text{open}}, T_{\text{close}}] \\
0, & \text{otherwise}
\end{cases}
\end{equation}

\subsubsection{Platform Root-of-Trust Attestation ($I_r$)}
Cryptographic attestation and hardware keystore validity are verified via the platform integrity predicate $\Phi_{\text{hw}}(d)$:
\begin{equation}
I_{r}(R) = \begin{cases}
1, & \text{if } \Phi_{\text{hw}}(d) = \text{TRUE} \\
0, & \text{otherwise}
\end{cases}
\end{equation}

\subsection{Deterministic Decision Function}
The composite decision engine maps each transaction request deterministically to a discrete outcome $\Omega = \{\text{ALLOW}, \text{REVIEW}, \text{BLOCK}\}$. The function signature is formalized strictly as $\mathcal{D}: \mathcal{T}_{\text{req}} \to \Omega$, with invariant vector $\mathcal{I}(R)$ evaluated internally over request $R$:
\begin{equation}
\mathcal{D}(R) = \begin{cases}
\text{BLOCK}, & \text{if } I_{a}(R)=0 \vee I_{m}(R)=0 \\
              & \quad \vee I_{t}(R)=0 \vee I_{r}(R)=0 \\
\text{REVIEW}, & \text{if } I_{\text{auto}}(R)=0 \vee I_{s}(R)=0 \\
\text{ALLOW}, & \text{if } \bigwedge_{j \in \{s, \text{auto}, a, m, t, r\}} I_{j}(R) = 1
\end{cases}
\end{equation}

Operational precedence is evaluated top-down: fatal policy violations ($I_a, I_m, I_t, I_r$) immediately yield \texttt{BLOCK}. In their absence, supervisory escalation conditions ($I_{\text{auto}}, I_s$) route the request to \texttt{REVIEW}. The transaction resolves to \texttt{ALLOW} if and only if all six invariants evaluate to $1$.

\subsection{The Architectural Non-Interference Invariant}
Downstream AI explanation is defined as a decoupled function $f_{\text{AI}}$ parameterized solely by post-decision state:
\begin{equation}
E = f_{\text{AI}}(X, \mathcal{D})
\end{equation}
where $X = (R, \mathcal{I}(R), \tau)$ represents the immutable execution context committed at ledger time $\tau$. To structurally isolate the financial ledger against prompt injection vulnerabilities~\cite{promptinjection2024}, FST Pay specifies the capability boundary:
\begin{equation}
\text{Cap}(f_{\text{AI}}) \cap \{\text{write-ledger}, \text{write-wallet}, \text{mutate-auth}\} = \emptyset
\end{equation}
Equation (11) specifies that explanation services hold no mutation rights. This isolation is enforced via read-only Kafka consumer credentials and separate database connection pools, ensuring that prompt outputs cannot alter authorization states $\mathcal{D}$.

% =========================================================================
% SECTION IV: SYSTEM ARCHITECTURE
% =========================================================================
\section{System Architecture}
\subsection{Two-Stage Structural Pipeline}
FST Pay enforces an architectural segregation between authorization authority and analytical explanation across two stages:
\begin{itemize}
    \item \textbf{Stage 1: Deterministic Safety-Gating (Pre-Authorization):} A low-latency evaluation service implemented as a native microservice. Incoming payment requests are evaluated concurrently across invariant vector $\mathcal{I}$. In-memory rule lookups execute in sub-millisecond time ($< 1\text{ ms}$), supporting an overall pre-authorization gateway budget target of $p_{99} < 15\text{ ms}$ without probabilistic dependencies on the critical rail.
    \item \textbf{Stage 2: Downstream AI Explanation (Post-Decision):} An out-of-band analytical service bound strictly to committed event streams. It consumes verified ledger records, parses contextual metadata, and invokes isolated LLM endpoints to synthesize pedagogical summaries and parental compliance reports.
\end{itemize}

\begin{figure*}[t]
\centering
\begin{tikzpicture}[node distance=0.45cm and 0.7cm, auto]
    % Tier 1: Clients
    \node [flowNode, minimum width=2.4cm] (ios) {iOS Native App\\(Swift / Keystore)};
    \node [flowNode, right=0.5cm of ios, minimum width=2.4cm] (android) {Android Native\\(Kotlin / Biometrics)};
    \node [flowNode, right=0.5cm of android, minimum width=2.4cm] (web) {React Web Portal\\(Merchant / Admin)};

    \begin{scope}[on background layer]
        \node [fit=(ios)(android)(web), draw=bwBlack, rounded corners=2pt, fill=bwLight, inner sep=4pt] (tier1) {};
        \node [above right, font=\tiny\bfseries] at (tier1.north west) {TIER 1: PRESENTATION \& ADAPTIVE CLIENT LAYER};
    \end{scope}

    % Tier 2: Gateway & Ingress
    \node [flowNode, below=0.55cm of android, minimum width=8.5cm, fill=bwWhite] (gateway) {
        \textbf{API Gateway \& Security Ingress (Envoy / Cloudflare WAF)} \\
        \scriptsize Transport Layer Security (TLS) 1.3 Termination \quad $\bullet$ \quad OAuth2/OIDC Token Validation \quad $\bullet$ \quad Rate Limiting \& Anti-DDoS
    };

    % Tier 3: Core Microservices
    \node [flowNode, below=0.55cm of gateway, xshift=-4.5cm, minimum width=2.4cm] (mUser) {User \& KYC\\Service};
    \node [flowNode, right=0.3cm of mUser, minimum width=2.4cm] (mWallet) {Double-Entry\\Ledger Service};
    \node [flowNode, right=0.3cm of mWallet, minimum width=2.4cm, fill=bwAlt] (mGate) {Deterministic Gating\\Rules Engine};
    \node [flowNode, right=0.3cm of mGate, minimum width=2.4cm] (mPay) {Payment Rail\\Orchestrator};
    \node [flowNode, right=0.3cm of mPay, minimum width=2.4cm, dashed] (mAI) {Downstream AI\\Explanation Worker};

    \begin{scope}[on background layer]
        \node [fit=(mUser)(mWallet)(mGate)(mPay)(mAI), draw=bwBlack, rounded corners=2pt, fill=bwLight, inner sep=4pt] (tier3) {};
        \node [above right, font=\tiny\bfseries] at (tier3.north west) {TIER 3: CORE BACKEND MICROSERVICES LAYER (SPRING BOOT \& PYTHON WORKERS)};
    \end{scope}

    % Tier 4: Data & Persistence Tier
    \node [flowNode, below=0.55cm of mGate, xshift=-3.5cm, minimum width=3.2cm] (dbAurora) {PostgreSQL Aurora\\(ACID Financial Ledger)};
    \node [flowNode, right=0.4cm of dbAurora, minimum width=3.2cm] (dbRedis) {Redis Cluster\\(Velocity Limit Caching)};
    \node [flowNode, right=0.4cm of dbRedis, minimum width=3.2cm] (dbKafka) {Apache Kafka Bus\\(Audit Event Streaming)~\cite{madaminov2024hybrid}};

    \begin{scope}[on background layer]
        \node [fit=(dbAurora)(dbRedis)(dbKafka), draw=bwBlack, rounded corners=2pt, fill=bwLight, inner sep=4pt] (tier4) {};
        \node [above right, font=\tiny\bfseries] at (tier4.north west) {TIER 4: DATA PERSISTENCE \& EVENT STREAMING INFRASTRUCTURE};
    \end{scope}

    % Tier 5: External Rails
    \node [flowNode, below=0.5cm of dbRedis, minimum width=11.5cm, fill=bwDark, text=bwWhite] (extRails) {
        \textbf{TIER 5: EXTERNAL BANKING, UPI 2.0 SWITCH \& ISSUING RAILS}~\cite{kumar2024growth} \\
        \scriptsize NPCI Unified Payments Interface \quad $\bullet$ \quad Card Networks (RuPay / Visa) \quad $\bullet$ \quad Mobile Push Notification Webhooks
    };

    % Inter-tier Connections
    \draw [line] (tier1.south) -- (gateway.north);
    \draw [line] (gateway.south) -- (tier3.north);
    \draw [line] (mWallet.south) |- (dbAurora.west);
    \draw [line] (mGate.south) -- (dbRedis.north);
    \draw [dashedLine] (mPay.south) |- (dbKafka.east);
    \draw [dashedLine] (dbKafka.north) -- (mAI.south);
    \draw [line] (tier4.south) -- (extRails.north);
\end{tikzpicture}
\caption{FST Pay: Complete multi-tier production software system architecture diagram.}
\label{fig:system_arch}
\end{figure*}
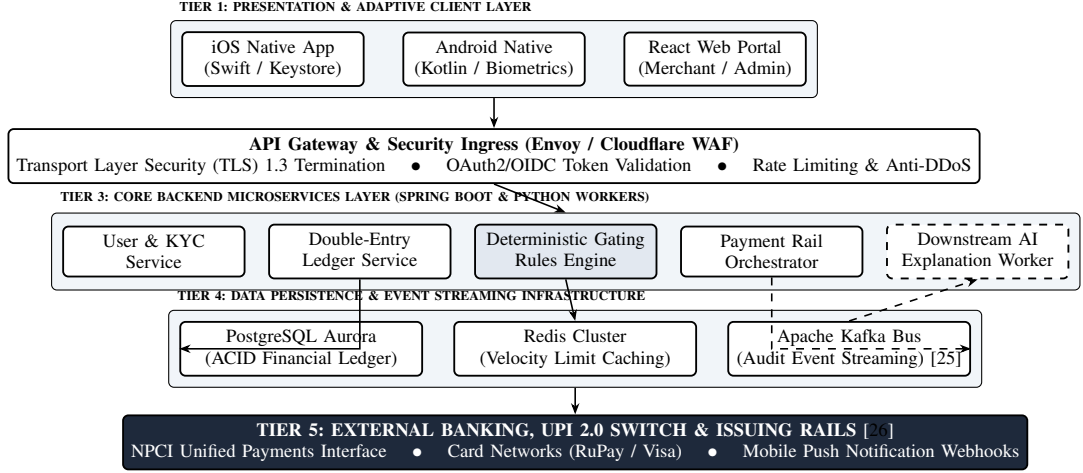

Figure~\ref{fig:system_arch} illustrates this multi-tier architecture from client presentation down to banking infrastructure.

% =========================================================================
% FIGURE 1: MULTI-TIER SYSTEM ARCHITECTURE DIAGRAM
% =========================================================================

\subsection{Component Decomposition}
In our reference software implementation, core transactional services are structured in Java using Spring Boot, with downstream workers deployed in Python:
\begin{enumerate}
    \item \textbf{Adolescent Client Application:} React Native mobile client maintaining encrypted local credential caches, biometric device authentication hooks, and budget visualization displays.
    \item \textbf{Guardian Supervisory Application:} Native companion interface dispatching push-notification webhooks for real-time transaction co-signing and limit adjustments.
    \item \textbf{API Gateway \& Ingress Service:} Enforces Transport Layer Security (TLS) 1.3 termination, rate-limiting, and short-lived JSON Web Token (JWT) signature verification.
    \item \textbf{Deterministic Rule Evaluator:} Stateless Spring Boot service maintaining in-memory Redis caches of active guardian policies and sliding-window velocity counters.
    \item \textbf{Core Payment Orchestrator:} Interfaces securely via ISO 20022 messaging structures (\texttt{pacs.008}) with underlying UPI Switch rails and National Financial Switch (NFS) endpoints~\cite{soundalgekar2024nfs}.
    \item \textbf{Downstream Generative Service:} Decoupled asynchronous worker service executing audited prompts against an isolated LLM endpoint to generate structured explainability logs.
\end{enumerate}

% =========================================================================
% SECTION V: PAYMENT DECISION AND AUTHORIZATION PIPELINE
% =========================================================================
\section{Payment Decision and Authorization Pipeline}
The authorization pipeline executes as a strictly ordered procedural gate prior to financial dispatch, guaranteeing predictable settlement without statistical ambiguity:
\begin{enumerate}
    \item \textbf{Syntactic Ingress \& Identity Validation:} The API gateway validates syntactic conformity, cryptographic payload signatures, replay protection nonces, and timestamp drift constraints ($|t - t_{\text{gateway}}| \le 5000\text{ ms}$). If the account state is inactive or suspended, execution halts immediately with a deterministic \texttt{BLOCK} response.
    \item \textbf{Velocity \& Cumulative Expenditure Assertion ($I_s, I_a$):} The engine evaluates cached ledger state to assert that transactional quantum $a$ respects both the single transaction cap $A_{\max}$ and the rolling window ceiling $L_{\Delta t}$. Fatal overages transition to \texttt{BLOCK}, whereas velocity-limit breaches trigger guardian escalation.
    \item \textbf{Category \& Curfew Enforcement ($I_m, I_t$):} Payee classifications are cross-checked against blacklisted Merchant Category Codes (e.g., MCC 7995 for gambling)~\cite{westermeier2024financial}, and current time is matched against curfew intervals $[T_{\text{open}}, T_{\text{close}}]$. Violations yield an immediate, non-overridable \texttt{BLOCK}.
    \item \textbf{Guardian Co-Sign Escalation Protocol ($I_{\text{auto}}$):} When the request exceeds the minor's autonomous limit $\theta_{\text{cosign}}$ or targets an unverified counterparty ($I_{\text{auto}} = 0$), the state machine transitions to \texttt{REVIEW}. A cryptographic push challenge is transmitted to the guardian's registered device. The transaction remains suspended in a pending authorization state until an authenticated biometric co-signature is received or an expiry threshold elapses.
    \item \textbf{Contextual Hardware Verification ($I_r$):} Device telemetry is evaluated against hardware-backed keystore integrity signals. If anomalous device manipulation is flagged, the pipeline halts with \texttt{BLOCK}.
    \item \textbf{Atomic Settlement Dispatch:} Upon satisfying all gating invariants (or receiving verified guardian consent), the orchestrator initiates settlement over interoperable UPI core switches and atomically updates the ledger.
\end{enumerate}

% =========================================================================
% SECTION VI: MULTI-PARTICIPANT PROCESS SEQUENCE
% =========================================================================
\section{Multi-Participant Process Sequence}
\begin{figure*}[t]
\centering
\begin{tikzpicture}[scale=0.85, every node/.style={transform shape}]
    \node [flowNode, minimum width=2.0cm] (teen) at (0,0) {Teen App\\User\_2317};
    \node [flowNode, minimum width=1.8cm] (gw) at (2.6,0) {API Gateway\\Edge Routing};
    \node [flowNode, minimum width=1.8cm] (rules) at (5.2,0) {Safety Gating\\Invariant Core};
    \node [flowNode, minimum width=1.8cm] (parent) at (7.8,0) {Guardian App\\Supervisor};
    \node [flowNode, minimum width=1.6cm] (rails) at (10.4,0) {UPI Rail\\NPCI Switch};
    \node [flowNode, minimum width=1.6cm] (ledger) at (12.8,0) {Aurora Ledger\\PostgreSQL DB};
    \node [flowNode, dashed, fill=bwLight, minimum width=1.8cm] (ai) at (15.5,0) {Downstream AI\\Explanation Agent};

    \draw [dashed, draw=gray] (teen) -- (0,-7.2);
    \draw [dashed, draw=gray] (gw) -- (2.6,-7.2);
    \draw [dashed, draw=gray] (rules) -- (5.2,-7.2);
    \draw [dashed, draw=gray] (parent) -- (7.8,-7.2);
    \draw [dashed, draw=gray] (rails) -- (10.4,-7.2);
    \draw [dashed, draw=gray] (ledger) -- (12.8,-7.2);
    \draw [dashed, draw=bwDark] (ai) -- (15.5,-7.2);

    \draw [line] (0,-0.8) -- node[above, font=\scriptsize] {1. Initiate Payment Request (\inr 1,200)} (2.6,-0.8);
    \draw [line] (2.6,-1.4) -- node[above, font=\scriptsize] {2. Validate Session \& Nonce} (5.2,-1.4);
    
    \draw [draw=bwBlack, fill=bwLight, line width=0.8pt] (5.0,-1.8) rectangle (5.4,-2.25);
    \node [right, font=\tiny\bfseries] at (5.5,-2.05) {3. Execute Invariant Rules};

    \draw [line] (5.2,-2.6) -- node[above, font=\scriptsize] {4. Co-Sign Approval Webhook} (7.8,-2.6);
    \draw [line] (7.8,-3.2) -- node[above, font=\scriptsize] {5. Guardian Co-Sign Response (Approved)} (5.2,-3.2);

    \draw [line] (5.2,-3.8) -- node[above, font=\scriptsize] {6. Authorized Payment Intent} (2.6,-3.8);
    \draw [line] (2.6,-4.4) -- node[above, font=\scriptsize] {7. UPI Settlement Request} (10.4,-4.4);
    \draw [line] (10.4,-5.0) -- node[above, font=\scriptsize] {8. Settlement Acknowledged (UTR)} (2.6,-5.0);

    \draw [line] (2.6,-5.6) -- node[above, font=\scriptsize] {9. Commit Settlement Record} (12.8,-5.6);

    % Boundary Line
    \draw [dashed, draw=bwBlack, line width=0.85pt] (-0.3,-6.1) -- (16.2,-6.1);
    \node [above, font=\tiny\bfseries] at (8.0,-6.1) {MANDATORY AUTHORIZATION BOUNDARY --- GENERATIVE AI CONSUMES AUDITED EVENTS ONLY};

    \draw [dashedLine] (12.8,-6.5) -- node[above, font=\scriptsize] {10. Publish Transaction Event} (15.5,-6.5);
    \draw [dashedLine] (15.5,-7.0) -- node[above, font=\scriptsize] {11. Dispatch AI Summary \& Educational Insight} (0,-7.0);
\end{tikzpicture}
\caption{Multi-participant sequence diagram illustrating deterministic payment clearance and downstream asynchronous explanation dispatch.}
\label{fig:sequence_diagram}
\end{figure*}
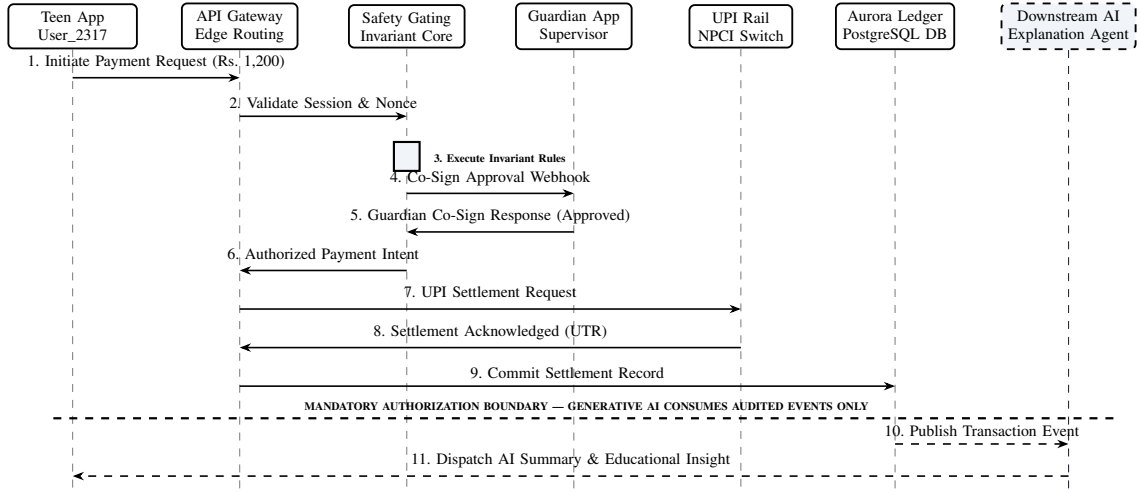

The end-to-end multi-participant interaction sequence is modeled in Figure~\ref{fig:sequence_diagram}. The primary actors and systemic entities include:
\begin{enumerate}
    \item \textbf{Adolescent User Application:} Authenticates the local user session, captures merchant destination parameters, and submits the cryptographically signed payment intent.
    \item \textbf{API Gateway:} Terminates ingress traffic, verifies mutual TLS (mTLS), and dispatches the sanitized intent to the deterministic engine.
    \item \textbf{Safety Gating Engine:} Concurrently evaluates invariant vector $\mathcal{I}$ against active in-memory state caches.
    \item \textbf{Guardian Application:} Receives synchronous push co-sign challenges for transactions requiring supervisory consent.
    \item \textbf{UPI Rail / Core Switch:} Executes fund transfers across participating payment service providers and commercial banks.
    \item \textbf{Aurora Ledger Core:} Records immutable double-entry journal entries under Atomicity, Consistency, Isolation, Durability (ACID) transactional isolation.
    \item \textbf{Downstream AI Worker:} Asynchronously consumes transaction finality events from Kafka queues and renders natural-language explanations.
\end{enumerate}

As demonstrated in Figure~\ref{fig:sequence_diagram}, rail settlement and ledger commitment occur strictly before event publication across the authorization boundary.

% =========================================================================
% FIGURE 2: SEQUENCE DIAGRAM TIKZ
% =========================================================================

% =========================================================================
% SECTION VII: DATA MODEL
% =========================================================================
\section{Data Model}
\begin{figure*}[t]
\centering
\begin{tikzpicture}[node distance=0.85cm and 1.1cm, auto, scale=0.82, every node/.style={transform shape}]
    % 1. USER
    \node [erTable] (user) {
        \begin{tabular}{p{3.8cm}}
        \rowcolor{bwDark}\color{white}\textbf{USER} \\
        \textbf{PK} \ \ user\_id: UUID \\
        \quad \ \ name: VARCHAR(100) \\
        \quad \ \ email: VARCHAR(120) \\
        \quad \ \ mobile\_number: VARCHAR(20) \\
        \quad \ \ role: VARCHAR(20) \\
        \quad \ \ account\_status: VARCHAR(20) \\
        \quad \ \ created\_at: TIMESTAMP \\
        \end{tabular}
    };

    % 2. WALLET
    \node [erTable, right=1.1cm of user] (wallet) {
        \begin{tabular}{p{3.8cm}}
        \rowcolor{bwDark}\color{white}\textbf{WALLET} \\
        \textbf{PK} \ \ wallet\_id: UUID \\
        \textbf{FK} \ \ user\_id: UUID \\
        \quad \ \ currency: VARCHAR(3) \\
        \quad \ \ current\_balance: NUMERIC(12,2) \\
        \quad \ \ locked\_balance: NUMERIC(12,2) \\
        \quad \ \ created\_at: TIMESTAMP \\
        \end{tabular}
    };

    % 3. TRANSACTION
    \node [erTable, right=1.1cm of wallet] (tx) {
        \begin{tabular}{p{4.0cm}}
        \rowcolor{bwDark}\color{white}\textbf{TRANSACTION} \\
        \textbf{PK} \ \ tx\_id: UUID \\
        \textbf{FK} \ \ wallet\_id: UUID \\
        \quad \ \ merchant\_id: VARCHAR(64) \\
        \quad \ \ amount\_inr: NUMERIC(10,2) \\
        \quad \ \ status: VARCHAR(20) \\
        \quad \ \ idempotency\_key: UUID \\
        \quad \ \ created\_at: TIMESTAMP \\
        \end{tabular}
    };

    % 4. APPROVAL_REQUEST
    \node [erTable, right=1.1cm of tx] (appr) {
        \begin{tabular}{p{4.0cm}}
        \rowcolor{bwDark}\color{white}\textbf{APPROVAL\_REQUEST} \\
        \textbf{PK} \ \ approval\_id: UUID \\
        \textbf{FK} \ \ tx\_id: UUID \\
        \textbf{FK} \ \ guardian\_id: UUID \\
        \quad \ \ status: VARCHAR(20) \\
        \quad \ \ resolved\_at: TIMESTAMP \\
        \quad \ \ created\_at: TIMESTAMP \\
        \end{tabular}
    };

    % 5. GUARDIAN_RELATIONSHIP
    \node [erTable, below=1.0cm of user] (rel) {
        \begin{tabular}{p{3.8cm}}
        \rowcolor{bwDark}\color{white}\textbf{GUARDIAN\_RELATIONSHIP} \\
        \textbf{PK} \ \ relationship\_id: UUID \\
        \textbf{FK} \ \ guardian\_id: UUID \\
        \textbf{FK} \ \ dependent\_id: UUID \\
        \quad \ \ is\_active: BOOLEAN \\
        \quad \ \ created\_at: TIMESTAMP \\
        \end{tabular}
    };

    % 6. SPENDING_LIMIT
    \node [erTable, below=1.0cm of wallet] (limits) {
        \begin{tabular}{p{3.8cm}}
        \rowcolor{bwDark}\color{white}\textbf{SPENDING\_LIMIT} \\
        \textbf{PK} \ \ limit\_id: UUID \\
        \textbf{FK} \ \ relationship\_id: UUID \\
        \quad \ \ daily\_limit\_inr: NUMERIC(10,2) \\
        \quad \ \ per\_transaction\_inr: NUMERIC(10,2) \\
        \quad \ \ category\_limit: JSONB \\
        \quad \ \ is\_active: BOOLEAN \\
        \end{tabular}
    };

    % 7. AI_EXPLANATION_LOG
    \node [erTable, below=1.0cm of tx] (aiLog) {
        \begin{tabular}{p{4.0cm}}
        \rowcolor{bwMid}\color{white}\textbf{AI\_EXPLANATION\_LOG} \\
        \textbf{PK} \ \ log\_id: UUID \\
        \textbf{FK} \ \ tx\_id: UUID \\
        \quad \ \ prompt\_tokens: INTEGER \\
        \quad \ \ completion\_tokens: INTEGER \\
        \quad \ \ latency\_ms: INTEGER \\
        \quad \ \ model\_name: VARCHAR(50) \\
        \quad \ \ created\_at: TIMESTAMP \\
        \end{tabular}
    };

    % 8. TRANSACTION_EXPLANATION
    \node [erTable, below=1.0cm of appr] (aiExp) {
        \begin{tabular}{p{4.0cm}}
        \rowcolor{bwMid}\color{white}\textbf{TRANSACTION\_EXPLANATION} \\
        \textbf{PK} \ \ explanation\_id: UUID \\
        \textbf{FK} \ \ tx\_id: UUID \\
        \quad \ \ explanation\_text: TEXT \\
        \quad \ \ model\_version: VARCHAR(30) \\
        \quad \ \ language: VARCHAR(10) \\
        \quad \ \ created\_at: TIMESTAMP \\
        \end{tabular}
    };

    % Relationships with non-colliding labels
    \draw [line] (user.south) -- (rel.north);
    \node [left=2pt, font=\tiny] at ($(user.south)!0.5!(rel.north)$) {1:N};

    \draw [line] (user.east) -- (wallet.west);
    \node [above=2pt, font=\tiny] at ($(user.east)!0.5!(wallet.west)$) {1:1};

    \draw [line] (wallet.east) -- (tx.west);
    \node [above=2pt, font=\tiny] at ($(wallet.east)!0.5!(tx.west)$) {1:N};

    \draw [line] (tx.east) -- (appr.west);
    \node [above=2pt, font=\tiny] at ($(tx.east)!0.5!(appr.west)$) {1:0..1};

    \draw [line] (rel.east) -- (limits.west);
    \node [above=2pt, font=\tiny] at ($(rel.east)!0.5!(limits.west)$) {1:1};

    \draw [dashedLine] (tx.south) -- (aiLog.north);
    \node [left=2pt, font=\tiny] at ($(tx.south)!0.5!(aiLog.north)$) {1:1};

    \draw [dashedLine] (tx.south) -- (aiExp.north west);
    \node [above=2pt, font=\tiny] at ($(tx.south)!0.5!(aiExp.north west)$) {1:1};
\end{tikzpicture}
\caption{Normalized relational database schema capturing core transactional invariants, guardian relationships, and decoupled AI explanation logs.}
\label{fig:database_schema}
\end{figure*}
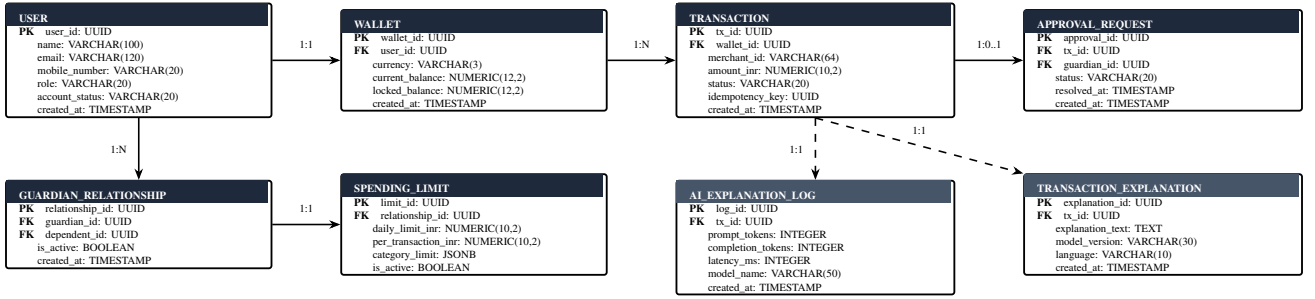

To support deterministic rule verification, low-latency audit logging, and isolated explanation ingestion, FST Pay employs the relational database schema illustrated in Figure~\ref{fig:database_schema}:
\begin{enumerate}
    \item \textbf{USER:} Stores master identities, distinguishing adolescent accounts (\texttt{ROLE\_TEEN}) from supervisory adult entities (\texttt{ROLE\_GUARDIAN}).
    \item \textbf{GUARDIAN\_RELATIONSHIP:} Represents legal guardian-dependent pairs, linking supervisory accounts to dependent wallets.
    \item \textbf{WALLET \& SPENDING\_LIMIT:} Decouples stored balances from dynamic limits, maintaining JSONB-encoded merchant restrictions and rolling ceilings.
    \item \textbf{TRANSACTION:} The immutable double-entry ledger table enforced with unique idempotency keys to prevent duplicate clearing under transient network retries~\cite{chandnani2024idempotency}.
    \item \textbf{APPROVAL\_REQUEST:} Records asynchronous co-sign states, resolution timestamps, and guardian cryptographic signatures.
    \item \textbf{AI\_EXPLANATION\_LOG \& TRANSACTION\_EXPLANATION:} Functionally isolated audit tables tracking prompt token usage, model identifiers, inferencing latencies, and generated natural-language explanations.
\end{enumerate}

% =========================================================================
% FIGURE 3: DATABASE SCHEMA TIKZ
% =========================================================================

\section{Implementation Considerations}
\subsection{Backend Stack and Microservice Orchestration}
The FST Pay reference architecture is organized into independent microservices using Java and Spring Boot for deterministic invariant evaluation, alongside Python services for asynchronous LLM orchestration:
\begin{itemize}
    \item \textbf{Deterministic Invariant Service:} Configured with embedded execution rules to achieve sub-millisecond in-memory evaluation. It maintains active user spending sums in a high-availability Redis cluster using sliding-window counter structures.
    \item \textbf{Payment Settlement Adapter:} Integrates with the UPI Switch via encrypted mutual TLS. It formats transaction requests following NPCI specifications and handles settlement responses.
    \item \textbf{Asynchronous Event Dispatcher:} Leverages an Apache Kafka event backbone. Once a transaction is settled or rejected, an event is emitted to a partitioned topic.
    \item \textbf{Downstream Generative Worker:} A decoupled Python service utilizing asynchronous workers to consume Kafka events, format prompts, and query language model endpoints.
\end{itemize}

\subsection{Idempotency and Concurrency Control}
Financial correctness requires strict concurrency control to prevent double-spending. When an adolescent client submits a payment request, the client generates a unique UUIDv4 idempotency key. The payment gateway verifies this key against the database using atomic insertion~\cite{chandnani2024idempotency}:
\begin{verbatim}
INSERT INTO transaction (idempotency_key, ...)
VALUES (?, ...) ON CONFLICT DO NOTHING;
\end{verbatim}
If a concurrent duplicate request arrives during authorization, it is intercepted and rejected immediately, ensuring that account balances are updated atomically without race conditions~\cite{chandnani2024idempotency}.

\section{Proposed Evaluation Framework and Theoretical Analysis}
\subsection{Evaluation Methodology}
In strict alignment with empirical reporting standards, no simulated or fabricated benchmark metrics are reported as proven experimental results. Instead, we formalize a rigorous proposed evaluation framework to validate FST Pay in future testbed implementations, as detailed in Table~\ref{tab:eval_framework}.

\begin{table}[h]
\centering
\caption{Proposed Evaluation Framework and Operational Metrics}
\label{tab:eval_framework}
\begin{tabular}{p{2.3cm}p{2.5cm}p{2.8cm}}
\toprule
\textbf{Evaluation Dimension} & \textbf{Target Metric} & \textbf{Verification Method} \\
\midrule
Deterministic Latency & Pipeline duration ($p_{99} < 15\text{ ms}$) & Ingress-to-egress Application Performance Monitoring (APM) distributed tracing \\
\addlinespace
Policy Correctness & Zero false allowances on restricted MCCs & Automated suite with combinatorial inputs \\
\addlinespace
Guardian Latency & Push-to-response duration ($t_{\text{cosign}}$) & Asynchronous round-trip event tracking \\
\addlinespace
Fault Isolation & System uptime under AI endpoint failure & Chaos engineering fault injection \\
\addlinespace
Audit Completeness & Ledger traceability ($100\%$ target) & Cryptographic ledger log verification \\
\bottomrule
\end{tabular}
\end{table}

\subsection{Theoretical Comparative Analysis}
Table~\ref{tab:comparative_analysis} presents a structured theoretical comparison contrasting FST Pay against traditional payment paradigms: Purely Probabilistic Fraud Detection, Unconstrained Generative AI Agents, and Conventional Static Retail Banking.

\begin{table*}[t]
\centering
\caption{Theoretical Comparison of Architectural Paradigms (Qualitative Design-Time Properties)}
\label{tab:comparative_analysis}
\begin{tabular}{p{3.2cm}p{3.2cm}p{3.2cm}p{3.2cm}p{3.2cm}}
\toprule
\textbf{Architectural Attribute} & \textbf{Purely Probabilistic ML} & \textbf{Generative AI Agent} & \textbf{Static Retail Banking} & \textbf{FST Pay (Proposed)} \\
\midrule
Decision Determinism & Low (probabilistic drift)~\cite{lu2024drift} & Nondeterministic (stochastic) & High (rigid Boolean rules) & \textbf{High (strict invariants)} \\
\addlinespace
Authorization Latency & Moderate (pipeline delay) & High (exceeds checkout SLA) & Minimal (static lookup) & \textbf{Design Target: $p_{99} < 15\text{ ms}$} \\
\addlinespace
Explainability Mechanism & Post-hoc (SHAP / LIME)~\cite{bauer2024expl} & Direct generative output & Error codes only & \textbf{Asynchronous Decoupled AI} \\
\addlinespace
Hallucination Hazard & Not applicable & Critical risk~\cite{hallucination2024fdm} & Zero & \textbf{Zero on Auth Rail (Isolated)} \\
\addlinespace
Guardian Oversight & Absent or manual & Unbounded & Binary account locks & \textbf{Granular Co-Sign Hold} \\
\addlinespace
Audit Integrity & Complex model weights & Non-reproducible & Standard database log & \textbf{Immutable Multi-Stage Log} \\
\bottomrule
\end{tabular}
\end{table*}

\section{Security and Privacy Considerations}
\subsection{Zero-Trust Ingress and Cryptographic Device Binding}
All API interactions require mutual TLS (mTLS) with public key pinning~\cite{diaz2024certpinning}. The adolescent mobile client establishes session authority using hardware-backed cryptographic keystores (e.g., Android Keystore, iOS Secure Enclave)~\cite{arm2020trustzone}. Transactions require biometric validation (fingerprint or facial recognition) combined with short-lived asymmetric authorization tokens~\cite{sethupathy2024evolving}.

\subsection{Isolation of the Generative Layer}
In compliance with NIST AI 600-1 recommendations for generative AI safety~\cite{nist2024genai}, the downstream explanation engine operates under least-privilege constraints~\cite{csa2023genai}. The service possesses read-only API access to the event queue. It has no network routes, database permissions, or API credentials capable of executing payment settlements, updating wallet balances, or altering authorization outcomes.

\subsection{Data Minimization and Privacy Preservation}
Adolescent user privacy is protected through strict data sanitization before event ingestion~\cite{opentelemetry2023scrubbing}:
\begin{itemize}
    \item Prompts dispatched to the downstream LLM contain zero personally identifiable information (PII). All user IDs, mobile numbers, and bank account numbers are tokenized or stripped~\cite{george2024tokenization}.
    \item Only broad contextual descriptors (e.g., transaction amount, merchant category, policy rule triggered) are provided to the explanation model.
    \item System logs are retained in encrypted cold storage with time-to-live (TTL) expiration schedules aligned with youth data privacy frameworks.
\end{itemize}

\section{Limitations and Future Work}
While FST Pay establishes a robust architectural framework, several operational limitations warrant discussion:
\begin{enumerate}
    \item \textbf{Absence of Production Benchmark Data:} As emphasized throughout this paper, FST Pay has not yet undergone wide-scale consumer field deployment. Empirical validation across diverse user cohorts remains future work.
    \item \textbf{Guardian Response Latency:} The REVIEW state depends on timely human parental intervention. If a guardian's device is offline or notifications are missed, transaction fulfillment will experience delays. Future work will investigate automated fallback policies and temporary micro-allowance overrides.
    \item \textbf{Evolving Merchant Categories:} Payee classification depends on accurate Merchant Category Codes (MCC). Malicious or deceptive merchants misrepresenting their category could bypass category-based invariants~\cite{su2024mccensemble}. Developing adaptive, verified merchant registries is an ongoing objective.
    \item \textbf{Lack of Machine-Checked Formal Verification:} The non-interference property is currently enforced via architectural capability partitioning. Machine-checked verification using static capability analyzers or automated model checkers is scoped for future work.
\end{enumerate}

\section{Conclusion}
This paper presented Financial Safety for Teens Pay (FST Pay), a digital payment safety architecture designed specifically for youth-oriented financial ecosystems. By establishing a clear architectural boundary between real-time deterministic safety gating and downstream generative explanation, FST Pay is designed to resolve the critical tradeoff between transaction safety and user-centric transparency.

The deterministic authorization engine evaluates payment requests across six comprehensive safety invariants, enforcing parentally configured spending limits, merchant restrictions, and guardian co-sign holds with sub-millisecond rule evaluation predictability. Meanwhile, the decoupled downstream generative AI engine receives finalized, read-only transaction events to generate plain-language explanations and financial literacy guidance without possessing authority over the payment ledger. This separation is architected to eliminate AI hallucination risks from financial clearance and to provide a structured foundation for future implementation and empirical evaluation in real-world digital payment platforms.

% =========================================================================
% REFERENCES (Clean IEEE Style with Clickable View Buttons ONLY)
% =========================================================================

\end{document}